\documentclass{aa}
\usepackage{graphicx}
\usepackage{psfig}
\usepackage{natbib}
\bibpunct{(}{)}{;}{a}{}{,}

\begin{document}

% our macros
\def\spose#1{\hbox to 0pt{#1\hss}}
\def\lta{\mathrel{\spose{\lower 3pt\hbox{$\mathchar"218$}}
     \raise 2.0pt\hbox{$\mathchar"13C$}}}
\def\gta{\mathrel{\spose{\lower 3pt\hbox{$\mathchar"218$}}
     \raise 2.0pt\hbox{$\mathchar"13E$}}}
\def\Msun{{\rm M}_\odot}
\def\msun{{\rm M}_\odot}
\def\Rsun{{\rm R}_\odot}
\def\Lsun{{\rm L}_\odot}
\def\half{{1\over2}}
\def\RL{R_{\rm L}}
\def\zs{\zeta_{s}}
\def\zR{\zeta_{\rm R}}
\def\dJJ{{\dot J\over J}}
\def\dMM{{\dot M_2\over M_2}}
\def\tKH{t_{\rm KH}}
\def\eck#1{\left\lbrack #1 \right\rbrack}
\def\rund#1{\left( #1 \right)}
\def\wave#1{\left\lbrace #1 \right\rbrace}
\def\dd{{\rm d}}
\def\new#1{{#1}}

\title{A transient variable 6 Hz QPO from GX~339-4}

\titlerunning{A transient variable 6 Hz QPO from GX~339-4}

\author{E. Nespoli\inst{1,2}
        \and
        T. Belloni\inst{1}
        \and
        J. Homan\inst{1}
        \and
        J.M. Miller\inst{3}
        \and
        W.H.G Lewin\inst{4}
        \and
        M. M\'endez\inst{5}
        \and
        M. van der Klis\inst{6}
}

\offprints{T. Belloni}

\institute{INAF -- Osservatorio Astronomico di Brera,
        Via E. Bianchi 46, I-23807 Merate (LC), Italy
   \and
        Universit\`a degli Studi di Milano,
        Via Celoria 16, I-20133, Milano, Italy
   \and
        Harvard-Smithsonian Center for Astrophysics, 60 Garden Street,
        Cambridge, MA 02138, USA
   \and
        Center for Space Research, Massachusetts Institute of Technology,
        77 Massachusetts Avenue, Cambridge, MA 02139-4307, USA
   \and
        SRON National Institute for Space Research, Sorbonnelaan 2,
        3584 CA Utrecht, the Netherlands
   \and
        Astronomical Institute ``Anton Pannekoek"
        University of Amsterdam and Center for High-Energy Astrophysics,
        Kruislaan 403, NL 1098 SJ Amsterdam, Netherlands.
}

\date{Received ???; accepted ???}

\abstract{ We report the results of an observation with the Rossi
X-ray Timing Explorer of the black hole candidate GX~339-4 during its
2002/2003 outburst. This observation took place during a spectral
transition from the hard to the soft state. A strong (6\% rms)
transient quasi-periodic oscillation (QPO) appears suddenly in the
power density spectrum during this observation. The QPO centroid is
$\sim$6 Hz, but it varies significantly between 5 and 7 Hz with a
characteristic time scale of $\sim$10 seconds, correlated with the
2-30 keV count rate. The appearance of the QPO is related to spectral
hardening of the flux, due to a change in the relative contribution of
the soft and hard spectral components. We compare this peculiar
behavior with results from other systems that show similar low
frequency QPO peaks, and discuss the results in terms of possible
theoretical models for QPO production.

}

\maketitle

\section{Introduction}

Variability studies of black-hole candidates (BHCs) have revealed the
presence of many types of quasi-periodic oscillations (QPOs) in the
X-ray flux of these systems \citep{va1995a,remumc2002b}. While most
BHC have shown QPOs between 0.01 and 30 Hz, in several of them also
QPOs above 40 Hz have been detected. In particular, QPOs between 1
and 10 Hz are very common in these systems and are probably linked to
similar features in neutron star systems \citep{bepsva2002}. As both
the low and high frequency QPOs are thought to arise in the accretion
flow close to the black hole, they are a potentially important tool in
determining the properties of these compact objects and the accretion
flow.

The presence and the properties of QPOs in BHCs seems to be related to
the spectral characteristics of the source. In this respect, transient BHCs are
particularly important for the study of QPOs, since these systems show
outbursts during which the mass accretion rate changes by several orders of
magnitude, usually resulting in several transitions between different
accretion modes or states. These states are characterized by distinct
spectral and temporal properties. QPOs are mostly found in states in
which the hard spectral component contributes significantly to the
spectrum and they are usually stronger at higher energies. Although a
considerable amount of data showing QPOs has been collected over the
years, many issues such as the nature of the frequency changes,
appearance/disappearance of the QPOs, relation to less coherent
variability, etc., have remained unsolved. Several theoretical models
exist for these oscillations, but there is no consensus as to their
physical nature \citep{alsh1985, stzhsw1996,milaps1998, stvi1998}.

GX~339-4, for which the  recently established mass function \citep[5.8 $\pm$0.5
$M_{\odot}$,][]{hystca2003} indicates the presence of a black-hole
primary, is usually referred to as a persistent BHC. However, its
long term behavior is more similar to that of transients and it is one
of the few sources that has shown all black hole spectral/timing
states \citep[see][for a detailed description of black hole
states]{tale1995,howiva2001,no2002}. While GX 339-4 is usually
observed in the low/hard state (LS), several state-transitions have
been reported \citep{mamama1984,mikiki1991,bemeva1999}.  Strong QPOs
were observed in the power density spectra of GX~339-4 during its
state transitions. \citet{mikiki1991} reported 1-15 Hz QPO together
with strong variable band-limited noise in the VHS transition observed
with Ginga. A complete analysis of similar oscillations in the
transient GS 1124-68 was performed by
\citet{tadomi1997}, who showed that there are two different sets of
QPOs, with different dependence on the source count rate. During these
observations, fast count rate variations were observed, coinciding
with changes in the spectral and timing properties
\citep{mikiki1991,tadomi1997}. So far, no high frequency QPOs have been
detected in GX 339-4.

After its 1998/1999 outburst, GX~339-4 returned to the LS, and it
became undetectable for the All-sky Monitor (ASM) onboard the Rossi
X-Ray Timing Explorer (RXTE) in May 1999, entering its `off' state,
observed before only once with EXOSAT \citep{ilch1986}.  The source
was detected with BeppoSAX on 1999 August 13, and its properties were
consistent with the off state being a low-luminosity extension of the
LS \citep{meva1997, kokuch2000}. However, a later observation with
BeppoSAX on 2000 September 10 revealed a flux level lower by a factor
of $\sim$4 than what was detected in 1999 \citep{conofe03}. GX~339-4 has
remained undetectable by the ASM until 2002. In March 2002, the source
became bright again \citep{smswhe2002} and we started an RXTE campaign
to follow its evolution. Here we present the results of the timing
analysis of a single observation from this campaign, in which a QPO
with exceptional behavior was observed. The full spectral/timing
analysis from all observations of our campaign will be presented
elsewhere. Preliminary results have been reported by
\citet{beneho2002}.

\section{Data analysis}

\begin{figure}[h] \resizebox{\hsize}{!}{\includegraphics{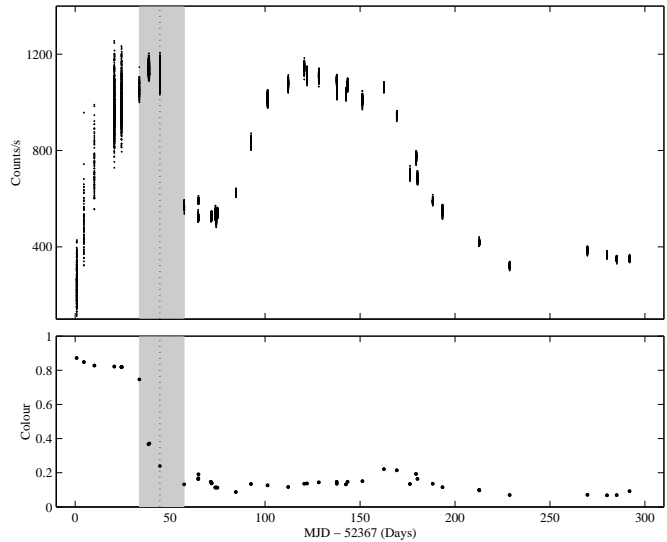}}
\caption{The 2.9-25.1 keV count rate (top) and color (bottom) evolution of
GX 339-4 during its 2002/2003 outburst, as observed with RXTE (PCU2
only).  The color is defined as the ratio of counts in the 16.4-19.8
keV band over counts in the 3.7-6.5 keV band. Day 0 corresponds to
April 3 2002 (MJD 52367), the vertical dotted line marks the time of
the observation analyzed here (May 17 2002/MJD 52411, see Fig.\ 2 for
blow up), and the gray area indicates the transition interval. The
time resolution of the light curve is 16s, while in the color curve
each point represents a single observation. Data are corrected for
background but not for dead time.}
\label{figure1_1}
\end{figure}

Figure \ref{figure1_1} shows the 2002/2003 outburst of GX 339-4. The
light and color curves are obtained from the data of the Proportional
Counter Array (PCA) onboard RXTE. Only counts from the most reliable
of the five units of the PCA (PCU2) were used. From the figure it is
evident that the outburst had a complex evolution, which can be
roughly divided in three parts. In the first part (day 0--34), the
count rate increased with time, showing a rather hard spectrum
gradually softening with time. The second part (day 34--57, depicted
by the gray strip) showed a drop in count rate accompanied by
considerable spectral softening. During the third part (day 57--292)
the count rate returned almost to the level of the first peak, after
which it showed an overall, but sometimes irregular, decay. These
changes occurred while the spectrum remained much softer than during
the first part of the outburst. From the light curve, it is also clear
that variability on a time scale of tens of seconds is much stronger in
the spectrally hard part of the outburst. Based on the timing and
color data, we conclude that the source evolved from the LS, through
the very high state (VHS) or intermediate state (IS), to the HS
\citep{beneho2002}.  A dotted line in Fig.\ \ref{figure1_1} marks the observation discussed in this work.

\begin{figure}[h] \resizebox{\hsize}{!}{\includegraphics{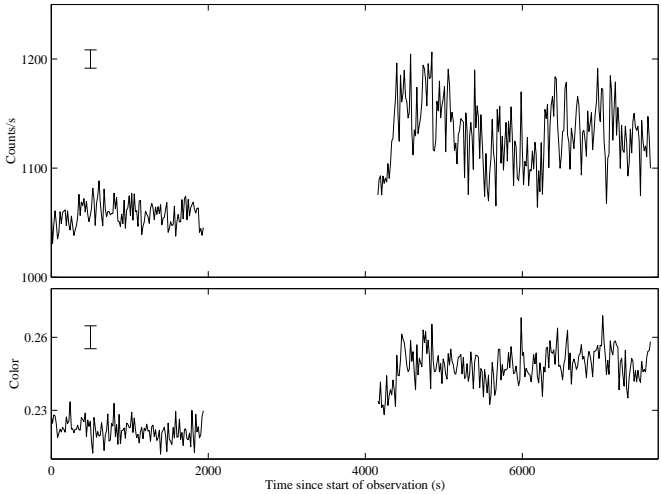}}
\caption{A blow up of the light and color curve for MJD 52411. Time
resolution is now 16s in both curves. Typical error bars are shown.
See caption of Fig.~1 for additional details.}
\label{figure1_2}
\end{figure}

This observation took place during the transitional phase of the
outburst and, although not outstanding in the light curve, showed
peculiar timing features, different from those observed before and
after. It corresponds to 2002 May 17/MJD 52411 (two RXTE orbits,
14:24-14:58 UT and 15:34-16:33 UT), with a total PCA exposure time of
5.58 ks. In Fig.\ \ref{figure1_2} we show a 16-s time resolution light
curve and the color evolution during this observation. Below, we
present the results of the timing and spectral analysis from these
data.

\subsection{Power Density Spectra}

\begin{figure}[h] \resizebox{\hsize}{!}{\includegraphics{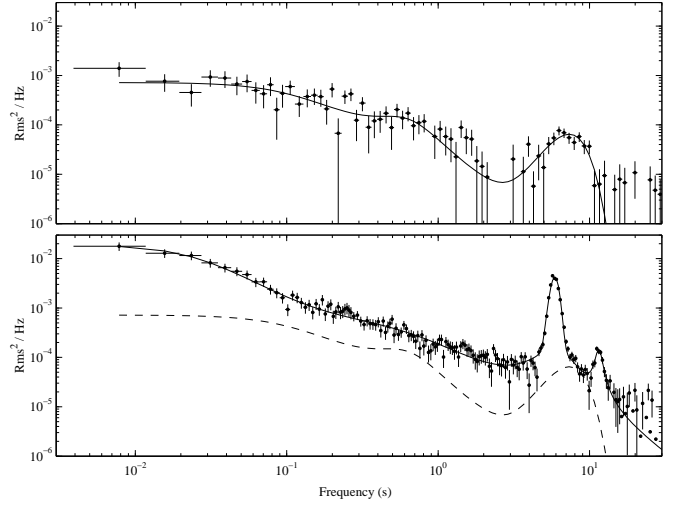}}
\caption{Power Density Spectra from the two RXTE orbits (top:
first orbit; bottom: second orbit).  The solid lines correspond to the
best-fit models described in the text. For reference, the dashed line
in the lower panel shows the best fit of the first orbit.}
\label{figure2} \end{figure}

For the timing analysis, we used {\tt Single bit} data from all active
PCUs, with a time resolution of 1.25$\times 10^{-8}$s and covering the
PCA channel range 0--35, corresponding roughly to 2--16 keV. For each
of the two orbits, we subdivided the data into stretches 128 seconds
long and produced a power density spectrum (PDS) for each of them.
The resulting PDS were then averaged, a logarithmic rebinning was
applied, the contribution due to Poissonian statistic was subtracted
\citep{zh1995,zhjasw1995}, and finally they were converted to
fractional squared rms \citep{beha1990}. The resulting power density
spectra for each orbit are shown in Fig.\ \ref{figure2}.
%No significant power is observed above 30 Hz.
The two PDS are markedly different. In the first orbit, the power density 
spectrum shows weak power-law noise with a very broad bump between 6
and 10 Hz. In the second orbit, the power-law noise is enhanced and a
strong QPO at $\sim$6 Hz appears, together with its second
harmonic. We fitted both power density spectra with a model consisting
of Lorentzian components \citep{mikiki1991, bepsva2002}. During the first orbit,
the broad QPO (centroid frequency 7.3$\pm$0.2 Hz and full-width at
half-maximum 4.4$\pm$0.4 Hz) and the continuum noise (consisting of
two Lorentzian components) have a fractional rms of 1.73$\pm$0.09\%
and 1.92$\pm$0.11\% respectively. A Lorentzian function proved to be a
poor approximation for the strong QPO in the second orbit. This effect
has also been observed in a QPO at a similar frequency in XTE
J1550-564 (Homan et al. 2001). A good fit to the QPO peak was obtained
with a model consisting of a Gaussian plus a broad Lorentzian to
account for the extended wings. The centroid
frequency of the Gaussian+Lorentzian complex is 5.895$\pm$0.004 Hz
(the values for the two components are constrained to be the
same). The width of the Gaussian is 0.351$\pm$0.006 Hz, and the total
(Gaussian plus Lorentzian) fractional rms amplitude is
6.68$\pm$0.07\%.  The continuum noise, fitted with two broad
Lorentzian components (as for the first orbit), has a total fractional
rms of 5.06$\pm$0.08\%. We repeated the analysis with data at higher
energies (PCA channels 36-51, corresponding to 17-24 keV). No signal
is detected during the first orbit, although with not very stringent
upper limits.  During the second orbit, the QPO fractional rms is
17.81$\pm$0.36\%.

\subsection{Time-frequency analysis}

In order to test whether the inadequacy of the Lorentzian model for
the  QPO is due to intrinsic variations of its centroid frequency, we
produced a spectrogram, defined (in its continuous version) as

\begin{equation}
S_x (t,\nu) =\left|{\int_{-\infty}^{+\infty} x(u) h^\ast (u-t)
  e^{-i2\pi \nu u} du}\right|^2 $$
\end{equation}

\begin{figure}[t] \resizebox{\hsize}{!}{\includegraphics{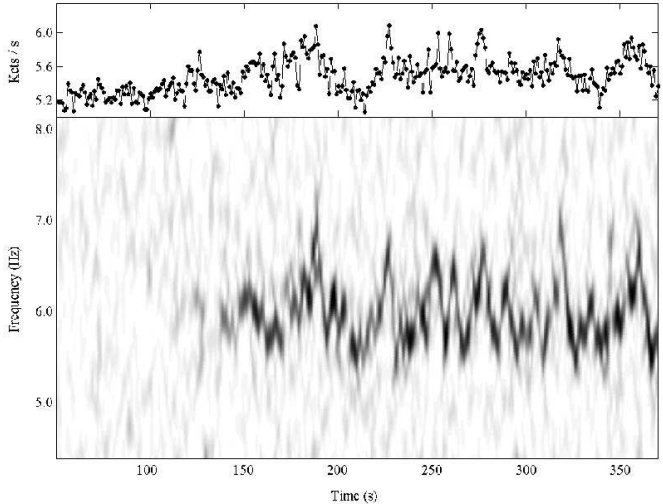}}
\caption{Top panel: PCA light curve (1 second binning)
50-370 seconds into the second RXTE orbit. Bottom panel: corresponding
spectrogram in the 4-8 Hz range; darker regions correspond to higher
power. } \label{figure3} \end{figure}

Here, $x(t)$ is our signal (PCA channel range 0-35, with a resolution
of 128$^{-1}$s) and $h(t)$ is a window function \citep[in our case, a
Welch window of length 4 seconds, see][]{prteve1992}. The time step
$t$, which sets the time resolution of the resulting time-frequency image
$S_x (t,\nu)$ was chosen to be 1 second. In order to suppress the
low-frequency noise, before applying Eqn. 1 we detrended our signal by
subtracting from it a spline fit to its 1-s rebinned
version. A section of the spectrogram, showing 320 seconds (starting
from 50 seconds into the second RXTE orbit) in the frequency range 4-8
Hz is shown in Fig.\ \ref{figure3}, together with the corresponding
light curve (in 1-second bins). Four effects can be observed:
\begin{itemize}

\item The QPO is not present in the first 120 seconds of the
observation, (we can set a 95\% confidence upper limit of 2.1\% rms by
fitting the PDS of the first 120 s with a QPO with centroid and width
fixed to the values reported above), barely visible in the next 20
seconds, and appears clearly only after that (to remain for the rest
of the RXTE orbit).

\item The light curve shows a net increase in count rate in
        correspondence with the onset of the oscillation, when also
        the enhanced  low-frequency noise, visible in Fig.\
        \ref{figure2}, turns on.

\item The centroid frequency of the QPO shows significant variations in
the range 5.5-7.0 Hz.

\item The centroid frequency shows a positive correlation with the
count rate (shown in the top panel).  \end{itemize}

\noindent On the basis of third point we conclude that the shape of the QPO, in
particular its broad wings, are indeed the result of the variations in the centroid frequency.
 
To examine in more detail the QPO-count rate relation, we fitted each
power density spectrum in the spectrogram (in the 4-8 Hz range) with a
model consisting of a Lorentzian peak and a power law for the local
continuum. Visual inspection of the results showed that the model was
able to recover accurately the local centroid frequency of the QPO.
On average, the FWHM of the best fit Lorentzian was found to be
$\sim$0.2 Hz. A correlation between QPO frequency and count rate over
the entire orbit is clear, although it shows a large scatter.
%A cross-correlation between the two time series (count rate and QPO
%centroid frequency) shows that the QPO frequency lags the count rate
%by about 1 second.

\begin{figure}[t]
\resizebox{\hsize}{!}{\includegraphics{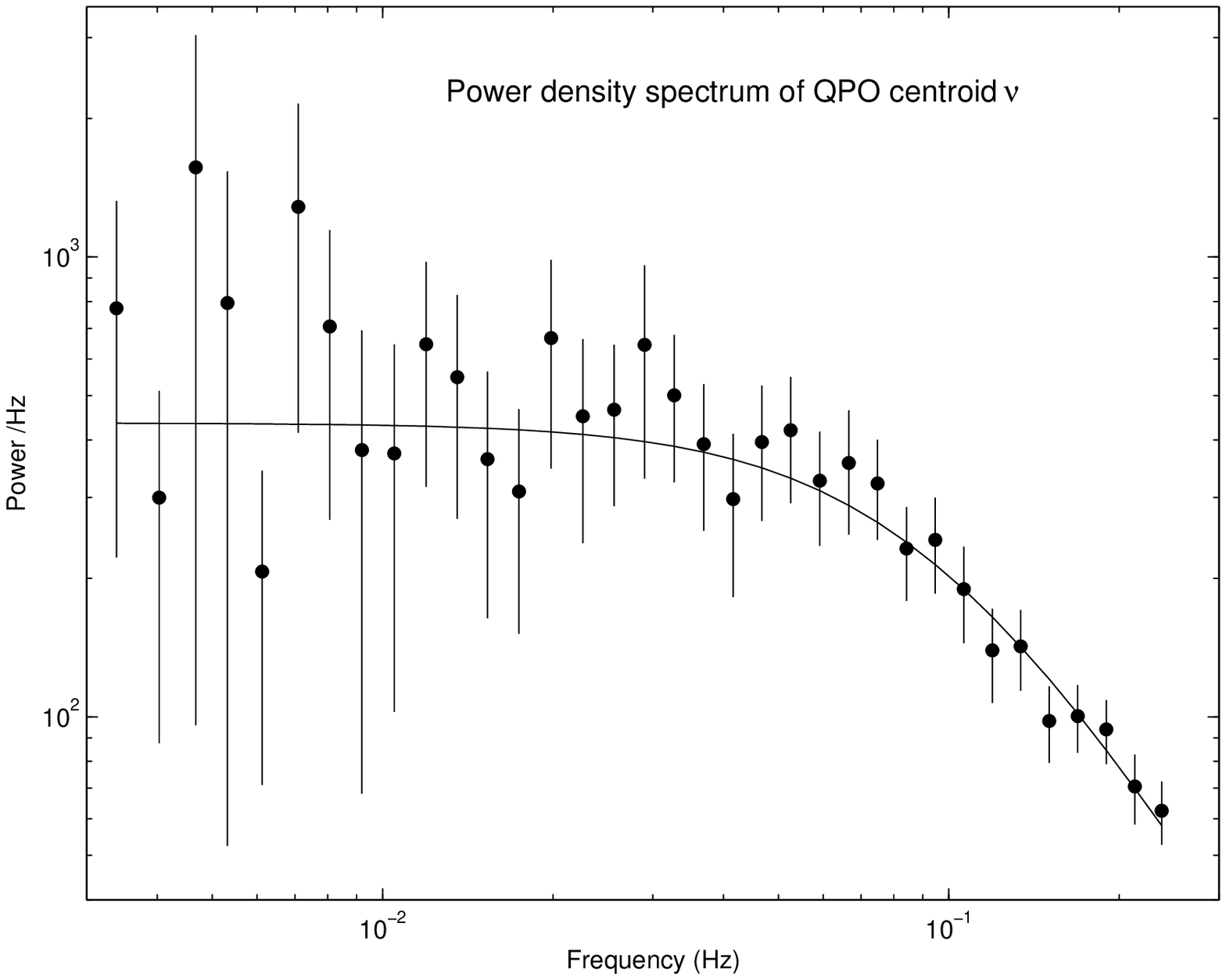}}
\caption{Power Density Spectrum of the QPO centroid frequency (see
text). Power is in units of Hz$^{2}$}. The line is a best fit with a zero-centered Lorentzian model.
\label{figure5}
\end{figure}

Since the QPO centroid frequency varies in an irregular way, we produced a
PDS of the QPO-frequency time series. In order to avoid problems due to the
fact that adjacent Welch windows overlap with each other, introducing
a net filtering effect, we repeated the whole analysis with a window
length of 2 seconds and non-overlapping windows. The  resulting power
density spectrum is shown in Fig.\ \ref{figure5}. A break in the power density 
spectrum is evident. A fit with a zero-centered Lorentzian yields a
characteristic frequency (HWHM) of 0.09 Hz. Fig.\ \ref{figure6} shows
the correlation between QPO frequency and total count rate from these
2-second windowed data. One can see that there is indeed a
correlation between the two quantities, but it is not very strict,
with a linear correlation coefficient of $\sim$0.5. Interestingly,
the additional noise component below 0.1 Hz in the second orbit,
reflects similar time scales as the variations of
the QPO frequency.

Although variability properties, such as QPO frequencies, generally
correlate better with spectral hardness than with count rate, the
statistical quality of our data prevented us from performing a
meaningful study with a color instead of count rate.

\subsection{Energy spectra}

We performed a spectral analysis of the two individual orbits separately,
ignoring the first 120 s of the second one. Spectra were created from
PCU2 data (2.9--25.2 keV) and HEXTE cluster A data (18.3--150 keV),
using the latest version of FTOOLS (5.2).  The spectra were corrected
for background and dead time effects, and fitted with XSPEC 11.2 using
a model consisting of a cutoff power law ({\tt cutoff}), a disk black
body ({\tt diskbb}), an accretion disk line ({\tt laor}) and a smeared
edge ({\tt smedge}). The interstellar absorption $N_H$ was fixed to a
value of $5\,10^{21}$ atoms/cm$^2$ \citep{ilch1986}.  For both orbits,
good fits were obtained.  The fit parameters were consistent between
the two orbits (see Table 1).
The only differences were found in the
fluxes of the disk black body and of the cutoff power law:
while the first decreased by 9\%, the second increased by more than
30\%, resulting in an overall hardening of the spectrum.
The calculated fluxes in the \mbox{2 -- 100 keV} range for the two RXTE orbits
are reported in Table 1.

\begin{table}
\begin{center}
\caption{\small Best fit parameters from the spectral fitting of the
data from the two RXTE orbits. Errors correspond to a 1$\sigma$ confidence
range. The total unabsorbed 2--100 keV luminosity was calculated assuming a distance of 4$\pm$1 kpc \citep{zdzpou:98}.}
\begin{tabular}{lcc}
\hline
\hline
   &  Orbit 1 & Orbit 2 \\
\hline
PL Index             & 2.44$\pm$0.02        & 2.44$\pm$0.10 \\
E$_{c}$ (keV)        & 190$^{+10}_{-90}$    & 110$^{+70}_{-40}$ \\
T$_{in}$ (keV)       & 0.89$\pm$0.01        & 0.92$\pm$0.01 \\
E$_{line}$ (keV)     & 6.86$\pm$0.11        & 6.7$\pm$0.2  \\
F$_{Tot}$$^a$        & (2.06$\pm0.06)\times10^{-8}$  & (2.25$\pm0.17)\times10^{-8}$  \\
F$_{disk}$$^a$       & (1.30$\pm0.04)\times10^{-8}$  & (1.18$\pm0.02)\times10^{-8}$  \\
F$_{pow}$$^a$        & (7.71$\pm0.35)\times10^{-9}$  & (1.02$\pm0.03)\times10^{-8}$  \\
L$_{tot}$ (erg s$^{-1}$)     & (3.73$\pm1.86)\times10^{37}$  & (4.07$\pm2.05)\times10^{37}$  \\
\hline
\end{tabular}
\\$^a$ unabsorbed 2--100 keV flux (erg cm$^{-2}$ s$^{-1}$)
\end{center}
\end{table}

%The two orbits mainly differ in the net count rate, with a
%significant average increase from the first orbit to the second: 1106
%cts/s (PCA counts) plus 14.7 cts/s (HEXTE counts) for the first
%orbit, 1187 cts/s (PCA counts) plus 18.9 cts/s (HEXTE counts) for the
%second one. This is also evident from the light curve shown in Fig.
%3, where at the very beginning of the second orbit, the counts are
%clearly below the mean value of all the rest.

\section{Discussion}

We have discovered a strong transient 6 Hz QPO in GX~339--4 with some
unusual properties: a sharp turn-on in time and large frequency
variability on a typical time scale of $\sim$10 seconds. The QPO was
found in an observation that was performed during a transition from a
hard to a soft state, during which the spectral and variability properties were those typical for a Very High State. It is interesting to note that,
while the observation occurred during a transition from a hard to a
soft spectral state, the appearance of the QPO in our observation is
associated with the hardening of the energy spectrum. This indicates
that such a hard to soft transition is not a smooth process
throughout the outburst.

\begin{figure}[t]
\resizebox{\hsize}{!}{\includegraphics{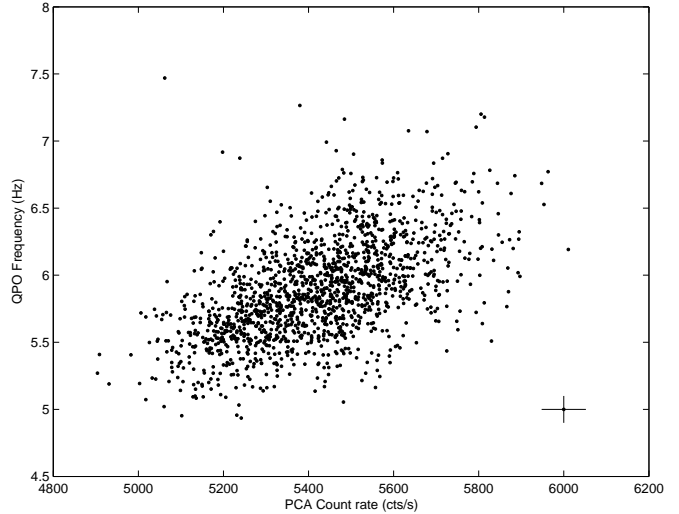}}
\caption{QPO frequency vs. PCA count rate for the 2-s binned data (see text). Typical error bars are shown. }
\label{figure6}
\end{figure}

The sharp 6 Hz QPO in the second orbit turned on within $\sim$20
seconds, similar to the timescale of the sharp increase observed in
the light curve. While a broad QPO was present during the first part
of our observation, it is not clear whether the 6 Hz QPO
evolved from this broad feature; the former is much weaker, less
coherent, and peaks at a different frequency. We note that in case
the broad feature is still present in the second part of the
observation, it might be partly responsible for the asymmetric wings
of the 6 Hz QPO (see Fig.\ \ref{figure2}). Such `shoulders' have been
reported before for similar 6 Hz QPOs in other BHCs. In the
terminology introduced by \citet{wihova1999} and \citet{resomu2002}
for QPOs in XTE J1550-564, the peak in the first orbit could be
interpreted as a type A QPO, and the ones in the second orbit as type
B; the observed spectral hardening is consistent with this
interpretation. Besides the changes around 6--7 Hz, also the
continuum at lower frequencies shows considerable changes. As noted
before, the increase in power below 0.1 Hz might reflect the count
rate variations that correlate with the changes in the QPO frequency
that have a similar time scale (see Fig.\ \ref{figure6}).

The observed FWHM of $\sim$0.2 Hz in the dynamical power spectrum is
consistent with being solely due to smearing by frequency changes.
This sets a lower limit to the Q value of the QPO of $\sim$30,
indicating a signal lifetime of at least 5 s. The fact that we see
uninterrupted frequency changes of 10 seconds and more suggests a
longer lifetime. If we associate the observed frequency with a certain
radius in the disk, it is unlikely that the QPO is caused by a single
blob of matter, as the blob would have to move inward and
outward. It is possible that what we observe is the result of
different blobs that appear and live at slightly different radii,
which vary on a similar time scale. However, Fig.\ \ref{figure3} and
Fig. 5 show that the QPO frequency variations are not consistent with
a white noise, i.e. there is a correlation between the QPO frequencies
at different times. This would mean that each new blob would ``know''
about the previous ones. In a model like that of
\citet{psno2001}, where a particular accretion disk radius acts as a
band-pass filter, the observed variations would translate into
corresponding variations of this filtering radius, resulting in a
natural correlation between different frequencies at different
times. The observed count-rate/frequency correlation would then
suggest a relation between this filtering radius and the X-ray flux.

Given its peculiarities, one might wonder whether this QPO detection
is unique.  A similar sharp QPO peak, needing a Gaussian model for a
satisfactory fit, was found by \citet{howiva2001} in the black-hole
transient XTE J1550-564. This QPO \citep[defined as type B
by][]{howiva2001} was also detected in a few observations with a
centroid frequency around 6 Hz. A sub-harmonic and two higher
harmonics were also detected. Interestingly, a similar sharp QPO was
observed in the Ginga data of the transient GS 1124-68
\citep{mikiig1994,bevale1997}. \citet{tadomi1997} classify the QPOs
observed in GS 1124-68 into two separate classes: the sharp QPO peak
was only observed in the range 5-7 Hz, similar to what we detected in
our data. \citet{howiva2001} also report the sudden appearance of a
QPO in XTE~J1550-564 associated to a count rate increase. However, in
this case the X-ray colors indicate a small and gradual softening of
the spectrum. A sharp QPO around 6 Hz, with harmonic and sub-harmonic
components and a similar low-frequency noise, was found by
\citet{cushha2000} in the transient XTE J1859+226. A narrow peak was
present in three of the four observations analyzed, with a centroid
frequency around 5.96 Hz, with properties similar to those seen in our
data.  Another narrow QPO peak was found in the bright transient
GRS~1739-27, although at 5 Hz
\citep{wimmj01}. The QPO, with its first overtone, was observed in the
second part of the data and not in the first part, where only a weak
noise component at low frequencies was noticed. \citet{wimmj01} found
that the frequency of the QPO increased together with the count rate,
with an even better correlation with the hard color. All these QPOs
have a centroid frequency of 5-7 Hz. It is not clear, however, why the
same frequency should play the same role in different systems. In all
these cases, the QPO is associated to a VHS-like energy spectrum, with
the presence of both a hard and a soft component, has a fractional rms
of a few percent and a FWHM similar to that shown here. In order not
to vary too much over systems with black holes of different masses,
the dependence of its frequency with mass of the central object must
be weak. 

The QPO observed here shows, moreover, characteristics in common with
both the HBO and the NBO of NS LMXB (see van der Klis 1995), letting
open the question if a stronger connection can be established either
with one or the other feature. HBOs present LFN (Low Frequency Noise),
with rms of 1-8 \%, the oscillation has a frequency that correlates
with count rate and an rms of 2-8 \%, the spectrum is hard. NBOs show
VLFN (Very Low Frequency Noise), with a rms of 0.6-4 \%, no LFN, the
frequency of the oscillation varies between 5 and 7 Hz, and its rms is
1-3 \%, while the energy spectrum is hard. Nothing can be said about
correlation between count rate and frequency over short time scale.

The rapid count-rate increase observed in our observation,
corresponding to a hardening of the flux, shows similarities to the
``dips'' and ``flip-flops'' reported by \citet{mikiki1991} from
GX~339-4 and by \citet{tadomi1997} from GS~1124-68, although the time
scale of our transition is slower. In both cases, GX~339-4 and GS
1124-68, a QPO peak was observed only in the high-flux intervals,
similar to our data, although the continuum noise was stronger at low
flux, contrary to what we observe here. The corresponding spectral
variations were similar to ours: the flux increase was caused by a
30\% increase of the hard component, while the disk component did not
vary. The spectral parameters remained unchanged \citep{mikiki1991}.
Notice that also in GS 1124-68, the centroid frequency of the sharp
QPO is positively correlated with count rate \citep{tadomi1997}.

The spectral hardening observed in connection to the appearance of
the QPO  suggests that the oscillation is associated to the hard
spectral component, as it is usually observed for low-frequency QPOs.
However, the presence of a soft component seems to be necessary for
the observation of such a QPO, as one can see also in the cases of
XTE J1550-564, GS 1124-68 and XTE 1859+226. A QPO with these
characteristics (narrow, around 6 Hz and with a weak noise component)
seems to be always associated to a sort of ``intermediate'' spectral
state, when both spectral components are present \citep[see
e.g.][]{ruleva1999}.

The results showed here pose constraints to theoretical models for the
production of the QPO. A successful model should be able to reproduce
a sharp appearance of a QPO like the one we observe, but also the
rapid variability of the characteristic frequency of the oscillation
(see Fig.\ \ref{figure5}) should be explained.

\begin{acknowledgements}

TB thanks the Cariplo Foundation for financial support. JH
acknowledges support from Cofin-2000 grant MM02C71842.

\end{acknowledgements}

\bibliographystyle{aa}
\bibliography{H4562}

\end{document}